\documentclass[11pt,twoside]{article}

\usepackage{asp2006}
\usepackage{epsf}
\usepackage{psfig}
\usepackage{lscape}

\usepackage{natbib}

\def\aap{A\&A}
\def\apj{ApJ}

\def\apjl{ApJ}
\def\mnras{MNRAS}

\def\nat{Nat}

\def\apjs{ApJS}
\newcommand{\beq}{
\begin{equation}
}
\newcommand{\eeq}{
\end{equation}
}
\newcommand{\kms}{\,{\rm km\,s^{-1}}}
\newcommand{\msun}{\,{\rm M_\odot}}

\def\spose#1{\hbox to 0pt{#1\hss}}
\newcommand{\lta}{\mathrel{\spose{\lower 3pt\hbox{$\mathchar"218$}}
     \raise 2.0pt\hbox{$\mathchar"13C$}}}
\newcommand{\gta}{\mathrel{\spose{\lower 3pt\hbox{$\mathchar"218$}}
     \raise 2.0pt\hbox{$\mathchar"13E$}}}
\def\simlt{\mathrel{\rlap{\lower 3pt\hbox{$\sim$}}\raise 2.0pt\hbox{$<$}}}
\def\simgt{\mathrel{\rlap{\lower 3pt\hbox{$\sim$}} \raise 2.0pt\hbox{$>$}}}

\begin{document}
\title{Evolution of massive black hole spins}   %%% Fill in title
\author{Marta Volonteri}   %%% Fill in author names
\affil{Department of Astronomy, University of Michigan, Ann Arbor, MI, 48109, USA}    %%% Fill in author affiliations

\begin{abstract} %%% Abstract to run on from here.
Black hole spins affect the efficiency of the ``classical" accretion processes, hence the radiative output from quasars. Spins also determine how much energy is extractable from the hole itself. Recently it became clear that massive black hole spins also affect the retention of black holes in galaxy, because of the impulsive ``gravitational recoil", up to thousands km/s, due to anisotropic emission of gravitational waves at merger. I discuss here the evolution of massive black hole spins along the cosmic history, due to the combination of  mergers and accretion events. I describe recent simulations of accreting black holes in merger remnants, and discuss the implication for the spins of black holes in quasars. 
\end{abstract}

\section{Introduction}
Black holes, as physical entities, span the full range of masses, from tiny black holes predicted by string theory, to monsters weighing by themselves almost as much as a dwarf galaxy (MBHs). Notwithstanding the several orders of magnitude difference between the smallest and the largest black hole known, all of them can be described by only three parameters: mass, spin and charge. Astrophysical black holes are even simpler, as charge can be neglected. So, besides their masses, $M_{\rm BH}$, astrophysical black holes are completely characterized by their dimensionless spin parameter, $a \equiv J_h/J_{max}=c \, J_h/G \, M_{\rm BH}^2$, where $J_h$ is the angular momentum of the black hole, and $0\le a\le 1$.  
%MBH spins influence the overall MBH growth and power production, as described in the following.  

\subsection{Radiative efficiency}
In radiatively efficient, geometrically thin accretion disks the mass-to-energy conversion efficiency, $\epsilon$, equals $\epsilon\equiv 1- E/M_{BH}c^2$, where $E$ is the binding energy per unit mass  of a particle in the innermost stable circular orbit (ISCO). The location of the ISCO depends solely on a black hole spin, shrinking by a factor of 6 between a non-rotating hole and its maximally rotating counterpart\footnote{We will use the term ``maximally rotating" for a black hole with $a=1$, although we note that \cite{Thorne1974} showed that accretion driven spin-up is limited to $a=0.998$.  Magnetic fields connecting material in the disk and the plunging region may further reduce the equilibrium spin. Magnetohydrodynamic simulations for a series of thick accretion disks suggest an asymptotic equilibrium spin at $a\approx 0.9$ \citep{Gammie2004}. The location of the ISCO depends also on the particle being on a prograde or retrograde orbit. The ISCO for a particle on a retrograde orbit around a hole with $a=1$ is 9 times larger than for its prograde counterpart.}. The closer the ISCO is to the horizon, the higher the mass-to-energy conversion efficiency, which increases from 6\% to 42\% in the above example. The mass-to-energy conversion  directly affects the mass-growth rate of black holes: high efficiency implies slow growth. More precisely, for a hole accreting at the Eddington rate, the black hole mass increases with time as:

\beq
M(t)=M(0)\,\exp\left({{1-\epsilon}\over{\epsilon}} \frac{t}{t_{\rm Edd}}\right), 
\eeq

 where $t_{\rm Edd}=0.45\,{\rm Gyr}$. The higher the spin, the higher $\epsilon$, implying longer timescales to grow the MBH mass by the same number of e--foldings. Going from $\epsilon=0.06$ to $\epsilon=0.42$, the difference in $M(t)$ amounts to 6 orders of magnitude, at $t=t_{\rm Edd}$. The typical spin therefore affects the overall mass-growth of MBHs, and  the duty cycle of quasars. 
\smallskip

The radiative efficiency is also the fundamental free parameter for the Soltan argument \citep{Soltan1982} and, more recently, synthesis models \citep[e.g.,][]{Merloni2008}  which relate the integrated MBH mass density to the integrated emissivity of  the AGN population, via the integral of the luminosity function  of quasars.
If the average efficiency of converting  accreted mass into luminosity is $\epsilon=L/\dot{M}c^2$, then the MBH will increase its mass by $\dot{M}_{BH}=(1-\epsilon)\dot{M}$, accounting for the fraction of the incoming mass that is radiated away. Applying this argument to the whole MBH population,  the MBH mass density can be related to the integral of the LF of quasar, $\Psi(L,z)$, with {\it the radiative efficiency being a free parameter.}
%

%\beq
%\rho_{BH}(z)=\int_z^\infty\frac{d t}{d z}d z
%\int_0^\infty \frac{(1-\epsilon)\,L}{\epsilon c^2}\Psi(L,z)d L.
%\eeq

%
%Recent results suggest that this approach might be simplistic, as the radiative efficiency evolves along the cosmic time \citep{Wang2009}. Quasars at the peak of their activity  ($z\sim2$) have high radiative efficiencies, hence have high spins. At later times ($z<1$)  the average radiative efficiency however decreases, hinting to lower spins. The theoretical investigation we propose here will address exactly this type of issues, by providing spin distributions of MBHs as a function of cosmic epoch. \\

\subsection{Relativistic jets} 

The so-called ``spin paradigm" asserts that powerful relativistic jets  are produced in AGNs with fast rotating
black holes \citep{Blandford1990}, implying that MBHs rotate slowly in radio-quiet quasars, which represent the majority of quasars \citep{Wilson1995}. 
However, if we expect the same mathematical and physical properties to describe both the black holes of a few solar masses and MBHs, such conjecture, at least in its basic interpretation, is at odds with studies of radio-emission from X-ray binaries \citep[e.g.,][]{Ulvestad2001,Kording2006}.  These works showed that the production of jets is intermittent, although the spin of stellar mass black holes is not expected to vary over the same timescales. {\it Why and when} jet form, and what is the role of spin (if any) in jet production has not been explained yet.

\subsection{Gravitational recoil} 

MBH spins affect the frequency of MBHs in galaxies, via the ``gravitational recoil" mechanism. When the members of a black hole binary coalesce, the center of mass of the coalescing system recoils due to the non-zero net linear momentum carried away by gravitational waves in the coalescence. If this recoil were sufficiently violent, the merged hole would breaks loose from the host and leave an empty nest.  Recent breakthroughs in numerical relativity have allowed reliable computations of black hole mergers and recoil velocities, taking the effects of spin into account. Non-spinning  MBHs, or binaries where MBH spins are {\it aligned} with the orbital angular momentum are expected to recoil with velocities below 200 $\rm{km\,s^{-1}}$. The recoil is much larger, up to thousands $\rm{km\,s^{-1}}$, for  MBHs with large spins in non-aligned configurations \citep{Campanelli2007,Gonzalez2007, Herrmann2007}. 

\section{Cosmic evolution of MBH spins}
MBH spins determine directly the radiative efficiency of quasars. On the other hand, accretion determines MBH spins. 
A hole that is initially non-rotating  gets spun up to a maximally-rotating state ($a=1$) after increasing its mass by a factor $\sqrt{6}\simeq 2.4$.    
A maximally-rotating hole is spun down by retrograde accretion to $a=0$ after growing by a factor $\sqrt{3/2}\simeq 1.22$. A $180^\circ$ flip of the spin of an extreme-Kerr hole will occur after tripling its mass.   Spin-up is therefore a natural consequence of prolonged disk-mode accretion: any hole that increases substantially its mass by capturing material with constant angular momentum axis would ends up spinning rapidly (``coherent accretion"). If the lifetime of quasars is long enough that angular momentum coupling between black holes and accretion discs through the Bardeen-Petterson effect effectively forces the innermost region of accretion discs to align, then quasar MBHs  should have large spins \cite{Volonterietal2005}. Spin-down occurs when counter-rotating material is accreted, if the angular momentum of the accretion disk is strongly misaligned with respect to the direction of the MBH spin.  If accretion proceeds via small (and short) uncorrelated episodes \citep[``chaotic accretion",][]{King2006}, where accretion of co-rotating and counter-rotating material is equally probable, then spins tend to be low.  This is because counter-rotating material spins MBHs down more efficiently than co-rotating material spins them up (as the ISCO for a retrograde orbit is at larger radii than for a prograde orbit, the transfer of angular momentum is more efficient in the former case.)  

MBH-MBH mergers also influence the spin evolution.  \cite{BertiVolonteri2008} consider how the dynamics of BH mergers influences the final spin. Except in the case of aligned mergers, a sequence of BH mergers can lead to large spins $ >0.9$ {\it only if} MBHs start already with large spins {\it and} they do not experience many major mergers.  Therefore, the common assumption that mergers between MBHs of similar mass always lead to large spins needs to be revised.

%compare three scenarios for the mass and spin co-evolution: (i) spins
%evolve only through mergers, (ii) spins evolve through mergers and prolonged
%accretion episodes, (iii) spins evolve through mergers and short-lived
%(chaotic) accretion episodes. If BHs accreted most of their mass through
%prolonged disc-mode accretion, by adding material with constant angular
%momentum axis, they would end up spinning rapidly. If instead BHs built-up
%their mass via short-lived episodes with uncorrelated angular momentum axis,
%the typical spin of BHs would be very low.

 \section{Spin evolution in gas-rich merger remnants}
Gas-rich mergers between galaxies of comparable mass (i.e. major mergers), are advocated to drive quasar activity, hence MBH growth.   The evolution of MBH spins has to be addressed by simulating the typical environmental conditions expected around quasars, where the properties of the accretion flow are established.  When accretion is triggered by galaxy mergers, as expected for quasars, the material that will feed MBHs is expected to assemble into a circumnuclear disk. These disks may be the end--product of gas--dynamical, gravitational torques excited during the merger, when large amounts of gas are driven into the core of the remnant  \citep{Mayer2007}.  This last phase of the merger has been suggested to be associated to  quasar activity. These are therefore the most relevant circumstances to be explored. During their inspiral MBHs are surrounded by a dense cocoon of gas that drives their dynamical decay and provides
fuel for the feeding of the holes \citep{Dottietal2006,Dotti2007}.  Since matter carries angular momentum also the spin vector can change during the accretion process.  The details of the dynamics may have a profound influence on the mass and spin evolution of the two MBHs.

\subsection{Simulation set-up}
\cite{Dotti2009b}  follow the dynamics of MBH pairs in nuclear discs using numerical
simulations run with the N--Body/SPH code GADGET (Springel, Yoshida \&
White 2001), upgraded to include the accretion physics \citep{Dotti2009a}. 
In our models, two MBHs are placed in the plane of a massive
circumnuclear gaseous disc, embedded in a larger stellar spheroid.
The disk is modeled with 2 $\times 10^6$
particles, has a total mass $10^8 \msun$, and follows a Mestel surface
density profile.  The spatial resolution of the hydrodynamical force
is $\approx 0.1$ pc. With this spatial resolution the Bondi radius of
the holes is resolved. The disk is rotationally supported and Toomre
stable. SPH disk particles evolve with a purely adiabatic equation of
state, described by either a polytropic index $\gamma= 7/5$ that
mimics a star-forming region \citep[``cold" disk,][]{Spaans00}, or $\gamma= 5/3$ that includes extra
heat, e.g, AGN feedback \citep[``hot" disk; see, e.g.][]{Mayeretal2006}.  
 
 The spheroidal component (bulge) is modeled with $10^5$ collisionless
particles, initially distributed as a Plummer sphere with a total mass
$\simeq 7\times$ the disc mass.  The mass of the bulge within $100$ pc
is five times the mass of the disc \citep{Downes1998}.  
 The two MBHs ($M_1$ and $M_2$) are equal in mass ($M_{\rm BH}=4\times
10^6\,\msun$). $M_1$, called primary for reference, is placed at rest at the centre of the
circumnuclear disc. $M_2$, termed secondary, is moving on an
initially eccentric ($e_0\simeq 0.7$) counterrotating (retrograde MBH) or corotating (prograde MBH) orbit with
respect to the circumnuclear disc. 
 Gas particles are accreted onto a MBH  if they lie within its Bondi radius,
and if the total mass accreted onto a MBH in a  timestep is lower than $\dot{M}_{Edd}$.  We find accretion rates close to Eddington for the central MBH ($M_1$) is in good agreement with the suggestion that quasars shine during the last phases of galaxy mergers. The secondary MBH in the simulation ($M_2$) accretes instead at sub-Eddington rates during most of its orbital decay.   

Each gas particle accreted by the MBH carries with it angular
momentum.  From the properties of the accreted particles we can
compute, as a function of time, the mass accretion rate and the versor
${\mathbf {\hat l}}_{\rm edge}$, that defines the direction of the total
angular momentum of the accreted particles. This information can be gathered only by
performing very high resolution simulations.  In our runs, the spatial resolution of the hydrodynamical force in the
highest density regions is $\approx 0.1$ pc, well below the Bondi radius. 

\subsection{Modelling the evolution of spin vectors}
We use the MBH accretion histories obtained from our SPH
simulations to follow the evolution of each MBH spin vector,
${\mathbf {J}}_{\rm BH}=(aGM_{\rm BH}^2/c)\hat{{\mathbf {J}}}_{\rm BH}$,
where $0 \leq a \leq 1$ is the adimensional spin parameter and
$\hat{\mathbf {J}}_{\rm BH}$ is the spin versor.  The scheme we adopt
to study the spin evolution is based on the model recently developed
by Perego et al. (2009). Here we summarize the algorithm used.

We assume that during any accretion event recorded in our SPH
simulations, the inflowing gas forms a geometrically thin/optically
thick $\alpha$-disc (Shakura \& Sunyaev 1973) on milli-parsec scales
(not resolved in the simulation), and that the outer disc orientation
is defined by the unit vector ${\mathbf {l}}_{\rm edge}.$ The evolution
of the $\alpha$-disc is related to the radial viscosity $\nu_1$ and
the vertical viscosity $\nu_2$: $\nu_1$ is the standard radial shear
viscosity while $\nu_2$ is the vertical shear viscosity associated to
the diffusion of vertical warps through the disc. The two viscosities
can be described in terms of two different dimensionless viscosity
parameters, $\alpha_1$ and $\alpha_2$, through the relations
$\nu_{1,2}=\alpha_{1,2}Hc_{\rm s}$, where $H$ is the disc vertical
scale height and $c_{\rm s}$ is the sound speed of the gas in the
accretion disc. We further assume $\alpha_2=f_2 / (2 \alpha_1)$, with
$\alpha_1 = 0.1$ and $f_2 = 0.6$ (Lodato \& Pringle 2007). We assume
power law profiles for the two viscosities, $\nu_{1,2} \propto
R^{3/4}$, as in the Shakura \& Sunyaev solution.

As shown by Bardeen \& Petterson (1975), if the orbital angular
momentum of the disc around the MBH is misaligned with respect to the
MBH spin, the coupled action of viscosity and relativistic
Lense-Thirring precession warps the disc in its innermost region forcing
the fluid to rotate in the equatorial plane of the spinning MBH.
The timescale of propagation of the warp is short comparted with
the viscous/accretion timescale so that
the deformed disc reaches an equilibrium profile that can be computed
by solving for the equation 

%\begin{eqnarray} \label{eqn:angular momentum}
%%\frac{\partial{\mathbf L}}{\partial t}= 
%\frac{1}{R}\frac{\partial}{\partial R}(R {\mathbf {L}} v_{\rm R})=
%\frac{1}{R}\frac{\partial}{\partial R}\left(\nu_1 \Sigma R^3 \frac{d\Omega}{dR}~ {\mathbf {\hat l}} \right)+ \nonumber \\
%% \qquad 
%+\frac{1}{R}\frac{\partial}{\partial R}\left(\frac{1}{2}\nu_2 R L \frac{\partial {\mathbf {\hat l}}}{\partial R} \right) 
%+ \frac{2G}{c^2} \frac{{\mathbf {J}}_{\rm BH} \times {\mathbf {L}}} {R^3} 
%\end{eqnarray}

\beq
\frac{1}{R}\frac{\partial}{\partial R}(R {\mathbf {L}} v_{\rm R})=
\frac{1}{R}\frac{\partial}{\partial R}\left(\nu_1 \Sigma R^3 \frac{d\Omega}{dR}~ {\mathbf {\hat l}} \right)+ 
\frac{1}{R}\frac{\partial}{\partial R}\left(\frac{1}{2}\nu_2 R L \frac{\partial {\mathbf {\hat l}}}{\partial R} \right) 
+ \frac{2G}{c^2} \frac{{\mathbf {J}}_{\rm BH} \times {\mathbf {L}}} {R^3} 
\label{eqn:angular momentum} 
\eeq

where $v_R$ is the radial drift velocity, $\Sigma$ is the surface
density, and $\Omega$ is the Keplerian velocity of the gas in the disc.
$\mathbf{L}$ is the local angular momentum surface density of the
disc, defined by its modulus $L$ and the versor ${\mathbf {\hat l}}$ that define its direction.

\begin{figure*}
\plotone{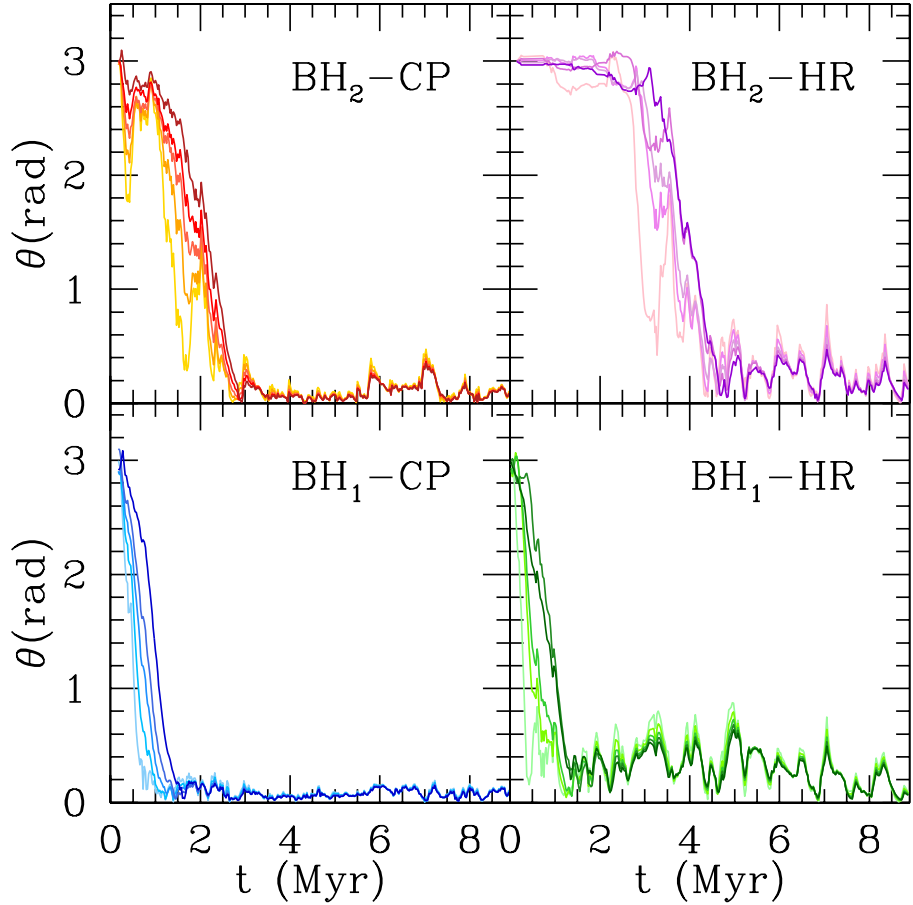}
\caption{Upper panels: time evolution of the relative angle between 
$M_2$ (secondary MBH) spin and the orbital angular momentum of the MBH pair. Left
(right) panel refers to a ``cold" (``hot") disk. The initial angle is arbitrarily
set to 2.5 radians (close to anti-aligned), and the initial spin
parameter magnitudes varies between 0.2 (lighter colours) to 1 (darker
colours).  Lower panels: same as upper panels for $M_1$ (primary MBH).  }
\label{fig:theta}
\end{figure*}

The boundary conditions to eq.~\ref{eqn:angular momentum} are the
direction of ${\mathbf {L}}$ at the outer edge ${\mathbf {\hat l}}_{\rm
edge}$, the mass accretion rate (that fixes the magnitude of
$\Sigma$), and the values of mass and spin of each MBH. All these
values but the MBH spins are directly obtained from the SPH runs.
In particular, the direction of the unit vector ${\mathbf {\hat l}}_{\rm
edge}$ is computed considering those SPH particles nearing the MBH
gravitational sphere of influence that are accreted according to the
criteria outlined above.

\begin{figure*}
\plottwo{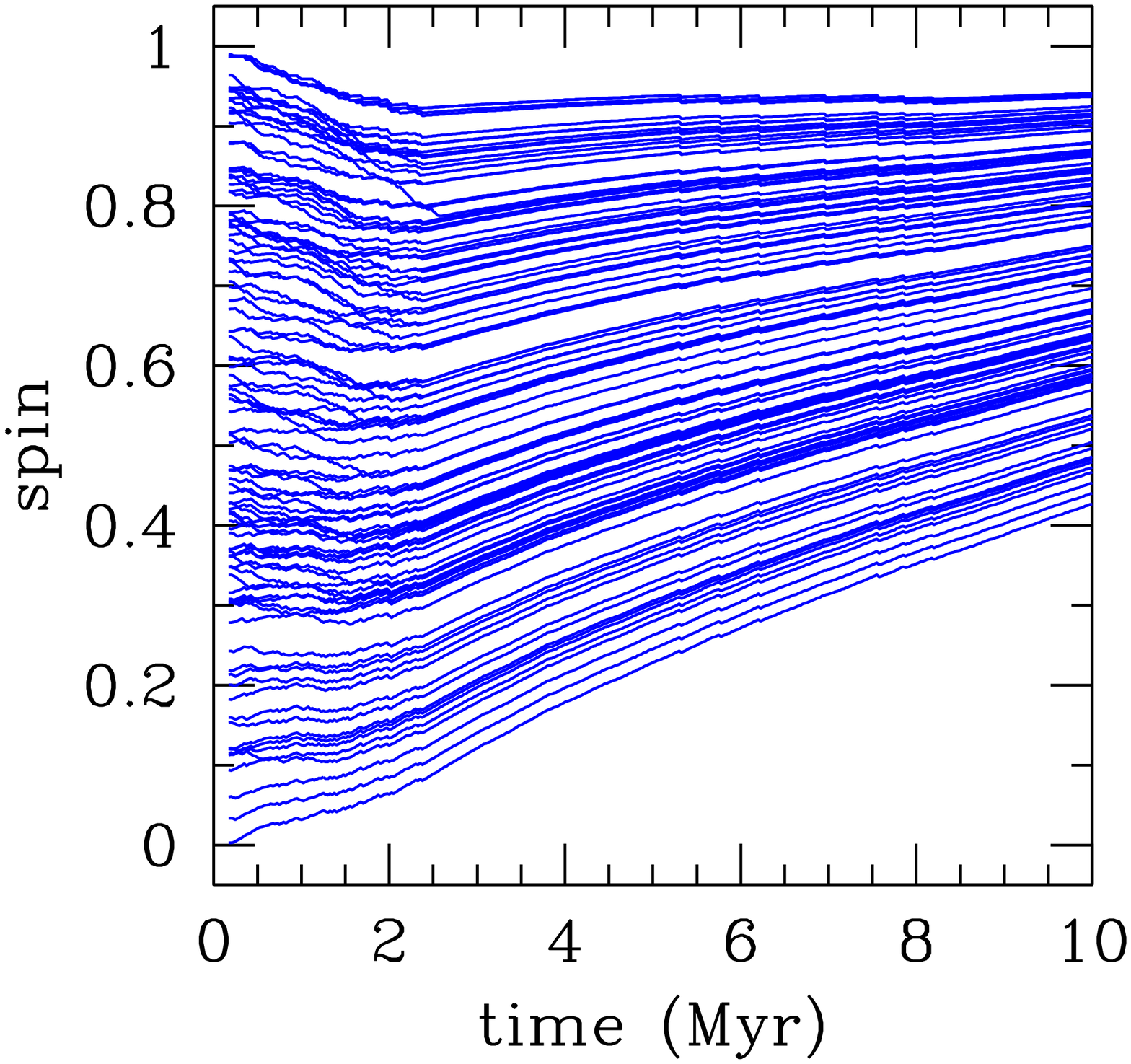}{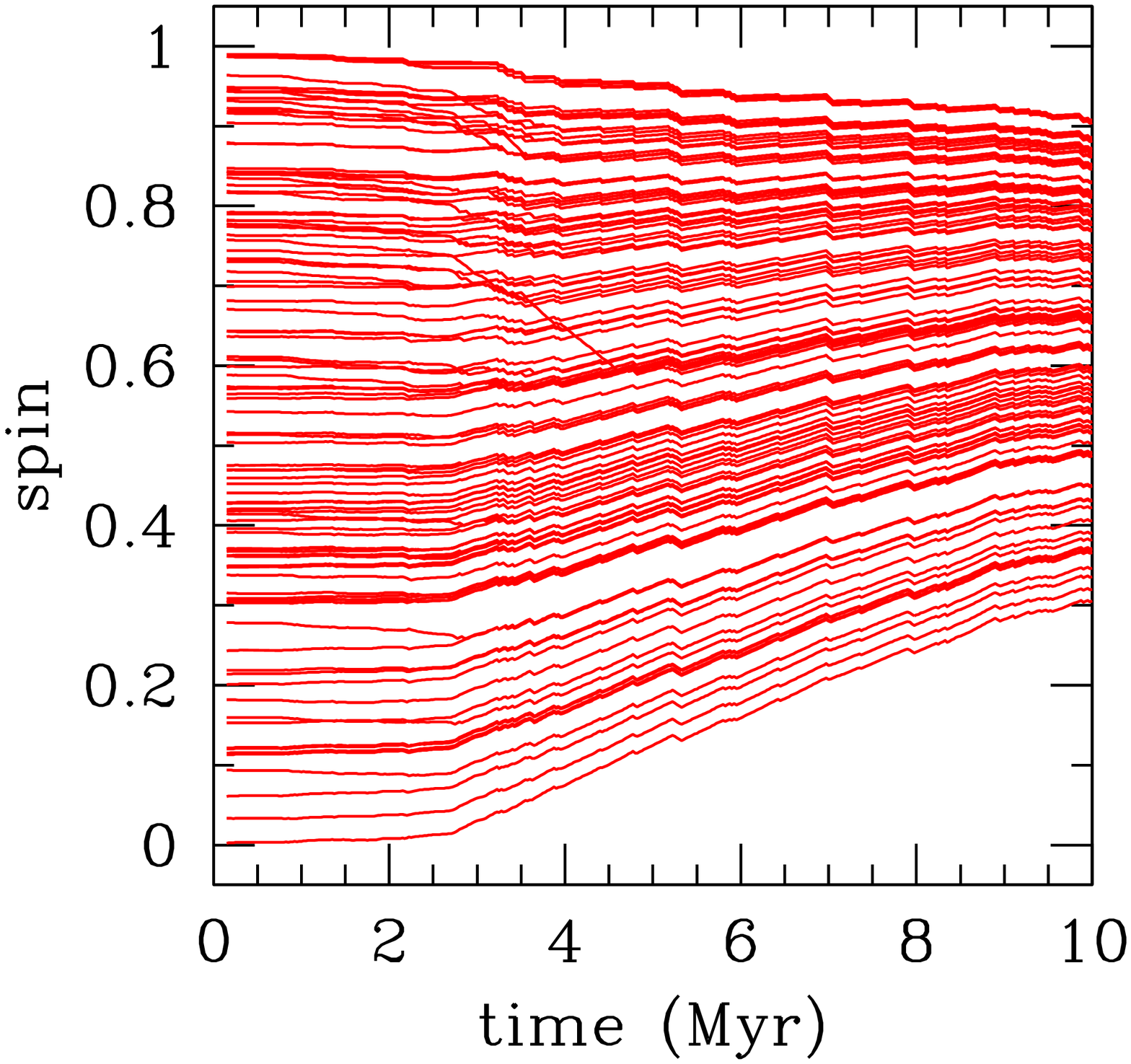}
\plottwo{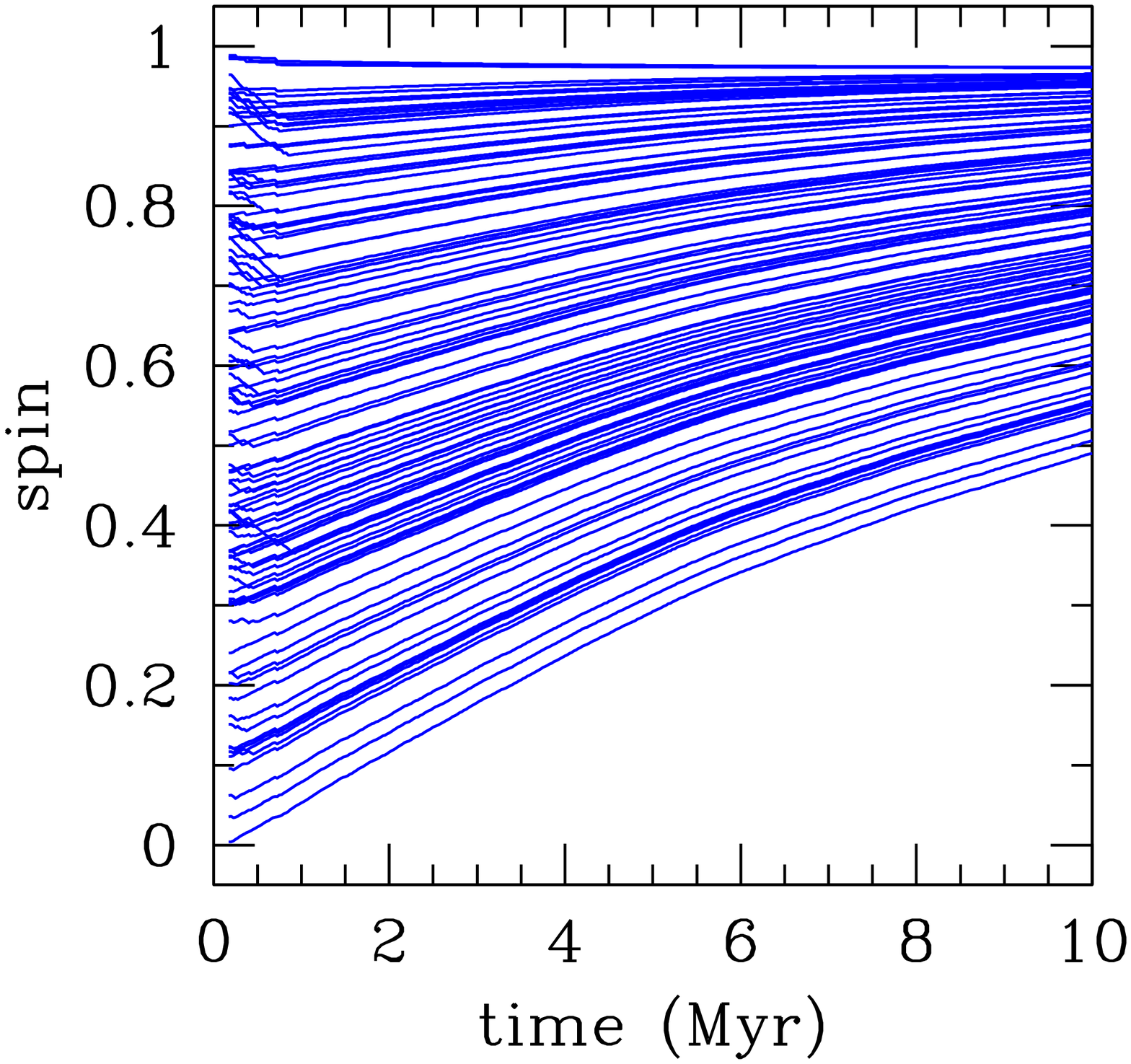}{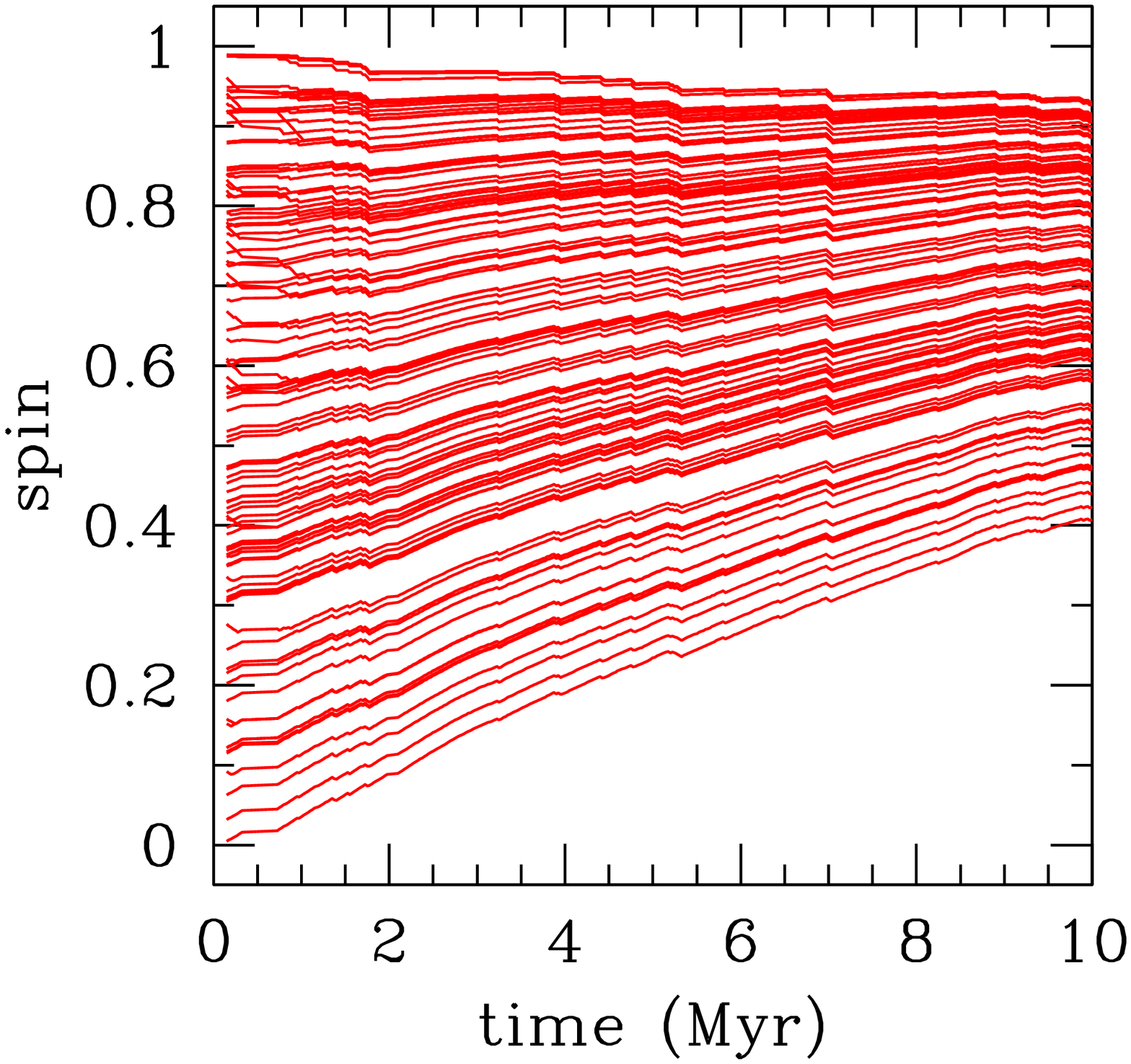}
    \caption{\footnotesize Left:  spin evolution for  MBHs embedded in a ``cold" circumnuclear disk. In a ``cold" disk, where turbulence and pressure are relatively unimportant, most of the accreted material has angular momentum directed along the same axis, leading to mostly ``coherent" accretion.  Right: same quantities, but for a MBH evolving in a more chaotic ``hot" circumnuclear disk. Here the direction of the angular momentum of the accreted gas is much more variable, giving rise to a ``chaotic" flow, and the magnitude of the MBH spins ends up being lower, at a given time along the simulation. Bottom: primary, central, MBH. Top: secondary MBH. In each panel we show 100 realizations starting from a flat distribution in spin magnitudes and initial orientations.     
                 \label{mdotti}}
\end{figure*}

The MBH spin changes also in
response to its gravito-magnetic interaction with the disc on a
timescale longer than the time scale of warp propagation (Perego et
al. 2009).  This interaction tends to reduce the degree of
misalignment between the disc and the spin decreasing with time the
angle between ${\mathbf {J}}_{\rm BH}$ and ${\mathbf {\hat l}}_{\rm
edge}$.  The MBH spin evolution is followed by solving the following equation:
\begin{equation} \label{eqn:jbh precession-disc}
\frac{d{\mathbf {J}}_{\rm BH}}{dt} =\dot{M}\Lambda(R_{\rm ISO})\hat
 {\mathbf {l}}(R_{\rm ISO}) + \frac{4\pi G}{c^2}\int_{\rm
 disc}\frac{{\mathbf {L}} \times {\mathbf {J}}_{\rm BH}}{R^2}dR.
 \end{equation}
 The first term in eq.~\ref{eqn:jbh precession-disc} accounts for the
 angular momentum deposited onto the MBH by the accreted particles at
 the innermost stable orbit (ISO), where $\Lambda(R_{\rm ISO})$
 denotes the specific angular momentum at $R_{\rm ISO}$ and $\hat
 {\mathbf {l}}(R_{\rm ISO})$ the unit vector parallel to ${\mathbf
 {J}}_{\rm BH}$, describing the warped disc according to the
 Bardeen-Petterson effect.  The second term instead accounts for the
 gravo-magnetic interaction of the MBH spin with the warped disc. It
 modifies only the MBH spin direction (and not its modulus), conserving
 the total angular momentum of the composite (MBH+disc) system (King
 et al. 2005).   We applied iteratively eq. 1 and 2 using inputs from the SPH
simulation that give the values of the mass accretion rate, the MBH
mass and the direction of ${\mathbf {\hat l}}_{\rm edge}$.  The
algorithm returns, as output, the spin vector, that is, its magnitude
and direction.  At each timestep our code therefore provides the angle
between the spin vector of each MBH and the angular momentum vector of
their relative orbit.

\subsection{Spins: alignment and magnitudes}
Figure~\ref{fig:theta} shows the time evolution of the relative angle
$\theta$ between the spin of each MBH and the orbital angular momentum
of the MBH pair. The initial relative angle has been arbitrarily set to 2.5
radians (143$^{\circ}$), while $a$ has initially five different values
(0.2, 0.4, 0.6, 0.8, and 1). In all cases we find that MBH spins loose memory of their initial
orientation: accretion torques suffice to align the spins with the
angular momentum of their orbit on a short timescale ($\simlt 1-2$
Myr). A residual off-set in the spin direction relative to the orbital
angular momentum remains, at the level of $\simlt 10^\circ$ for the case
of a cold disc, and $\simlt 30^\circ$ for a warmer disc. Alignment in a
cooler disc is more effective due to the higher coherence of the
accretion flow near each MBH that reflects the large-scale coherence of
the disc's rotation. If the MBHs coalesce preserving the spin
directions set after formation of a Keplerian binary, the MBH resulting
from their coalescence receives a relative small kick, $\simlt 100 \kms$.

Fig.~\ref{mdotti} shows the time evolution of the spin magnitudes (Dotti et al. in preparation). We show here 100 realizations starting from a flat distribution in spin magnitudes and initial orientations.   We find that initially maximally rotating are spun down, and initially slowly rotating holes are spun up, leading to intermediate equilibrium values ($\simeq 0.6-0.8$). The spin of the secondary MBH tends to be slightly lower, because of a higher incidence of counterrotating events.  It is suggestive that measured spins  cluster around these values.  The best measurement to date is the Seyfert galaxy MCG--6-30-15.  In this Seyfert, the iron line is so broad as to rigorously require $a=0.9\pm0.1$ \citep{Brenneman2006}. \cite{Miniutti2009} reported a spin parameter of $a = 0.6\pm 0.2$ in the narrow-line Seyfert-1 AGN SWIFT J2127.4$+$5654.  The X--ray spectrum of  Fairall~9 also results in a spin $a=0.7\pm0.2$ at 99\% confidence \citep{Schmoll2009}.

%\begin{figure}[t]
%\plottwo{procold_all.eps}{rethot_all.eps}
%    \caption{\footnotesize Left: accretion rate (top, in Eddington units), fraction of the angular momentum of the accreted material along the z-axis (middle), and spin (bottom) for  two MBHs embedded in a ``cold" circumnuclear disk (central MBH, $M_1$: solid curves; secondary MBH, $M_2$: dashed curves). In a ``cold" disk, where turbulence and pressure are relatively unimportant, most of the accreted material has angular momentum directed along the same axis, leading to mostly ``coherent" accretion.  The bottom panel shows the spin evolution for two different initial values of the spin: $a=0.2$ and $a=1$. Right: same quantities, but for MBHs evolving in a more chaotic ``hot" circumnuclear disk. Here the direction of the angular momentum of the accreted gas is much more variable, giving rise to a ``chaotic" flow, and the magnitude of the MBH spins ends up being lower, at a given time along the simulation.
%             \label{mdotti}}
%\end{figure}

\section{MBH spins and galaxy morphology}
If the events powering quasars coincide with the formation of elliptical galaxies (di Matteo et al. 2005), we might expect that
the MBH hosted by an elliptical galaxy had, as last major accretion
episode, a large increase in its mass. During this episode the spin increased significantly as well, possibly up to very high values, $a \simeq 0.6-0.8$, as suggested in the previous section. 
%Subsequently the black hole might have grown by swallowing the occasional molecular cloud, or by tidally disrupting stars. If the total contribution of these random episodes represents a small fraction of the BH mass, the spin is,
%however, kept at high values.

Black holes in spiral galaxies, on the other hand, probably had
their last major merger (i.e., last major accretion episode), if
any, at high redshift, so that enough time elapsed for the galaxy
disc to reform. Moreover, several observations suggest that single accretion
events last  $\simeq 10^5$ years in Seyfert galaxies, while the total activity lifetime (based on the fraction of disc galaxies that are Seyfert) is
$10^8-10^9$ years \citep[e.g.,][]{Kharb2006,Ho1997}. This
suggests that accretion events are very small and very
{`}compact'. Smaller MBHs, powering low luminosity AGN, likely grow by accreting smaller packets of material, such as
tidally disrupted stars \citep[for MBHs with mass $<2\times 10^6 \msun$,][]{
Milosavljevic2006}, or possibly molecular clouds \citep{HopkinsHernquist2006}.

Compact self-gravitating cores of molecular clouds (MC) can occasionally reach subparsec regions. Although the rate of such events is uncertain, we can adopt the estimates of Kharb et al. (2006), and assume that about $10^4$ of
such events happen within the total activity lifetime of a Seyfert. We can further assume a lognormal distribution (peaked at
$\log(M_{\rm MC}/\msun)=4$, with a dispersion of 0.75)
for the mass function of MC close to galaxy centers \citep[based on the
Milky Way case, e.g.,][]{Perets2007}. The result is, on the whole,
similar to that produced by minor mergers of black holes 
\citep{HB2003}, that is a spin down in a random walk fashion \citep{VSL07}. 

In a gas-poor elliptical galaxy, however, substantial populations of molecular clouds are lacking (e.g., Sage et al. 2007), eliminating this channel of MBH feeding. Main sequence stars, however, linger in galaxy centers. Tidal disruption of stars is a feeding mechanism that has been proposed long ago \citep{Hills1975, Rees1988}. One expects discs formed by stellar debris to form with a random orientation. Stellar disruptions would therefore contribute to the spin-down of MBHs. In an isothermal cusp, assuming that MBH masses scale with
the velocity dispersion, $\sigma$, of the galaxy (we adopt here the
Tremaine et al. 2002 scaling), we can derive the relative mass
increase for a MBH in 1 billion years:
\begin{equation} \frac{M_*}{M_{\rm
BH}}=0.37\left(\frac{M_{\rm BH}}{10^6\msun}\right)^{-9/8}.
\label{eqTD}
\end{equation}
The maximal level of spin down would occur assuming that all the tidal disruption events form counterrotating discs, leading to retrograde accretion. 
Eq. \ref{eqTD} shows that a small (say $10^5 \msun$) MBH starting at $ a=0.998$ would be spun down completely, as its mass increase is larger than $\sqrt{3/2}$ (cfr. section 2). On the other hand the spin of a larger (say $10^7 \msun$) MBH would not be changed
drastically. This feeding channel is likely efficient in early type discs which typically host faint bulges characterized by steep cusps, as exemplified above.  The situation is different for giant ellipticals: 
the central density profile displays a shallow core, and
tidal disruption of stars is unlikely.

\cite{VSL07} therefore suggest the spin of MBHs in giant elliptical galaxies is likely dominated by massive accretion events which follow gas--rich galaxy mergers (see Dotti et al. in this volume for a discussion of spin evolution in ellipticals formed in gas--poor mergers).  Both tidal disruption of stars, and accretion of gaseous clouds is unlikely in shallow, stellar dominated galaxy cores.  In a galaxy displaying instead power-law (cuspy) brightness profiles, the rate of stellar tidal disruptions is much higher and random small mass accretion events contribute to spin MBHs down.  This result is in agreement with Sikora et al. (2007) who found that disc galaxies tend to be weaker radio sources with respect to elliptical hosts. 

\acknowledgements %%% Text of acknowledgements runs on after this command.

M.V. acknowledges support from NASA award ATP NNX10AC84G. 

%\bibliographystyle{mn2e}
%\bibliography{paper}

\begin{thebibliography}{}

\bibitem[\protect\citeauthoryear{{Berti} \& {Volonteri}}{{Berti} \&
  {Volonteri}}{2008}]{BertiVolonteri2008}
{Berti} E.,  {Volonteri} M.,  2008, \apj, 684, 822

\bibitem[\protect\citeauthoryear{{Blandford}, {Netzer}, {Woltjer},
  {Courvoisier} \& {Mayor}}{{Blandford} et~al.}{1990}]{Blandford1990}
{Blandford} R.~D.,  {Netzer} H.,  {Woltjer} L.,  {Courvoisier} T.~J.-L.,
  {Mayor} M.,  eds, 1990, {Physical processes in active galactic nuclei.}

\bibitem[\protect\citeauthoryear{{Brenneman} \& {Reynolds}}{{Brenneman} \&
  {Reynolds}}{2006}]{Brenneman2006}
{Brenneman} L.~W.,  {Reynolds} C.~S.,  2006, \apj, 652, 1028

\bibitem[\protect\citeauthoryear{{Campanelli}, {Lousto}, {Zlochower} \&
  {Merritt}}{{Campanelli} et~al.}{2007}]{Campanelli2007}
{Campanelli} M.,  {Lousto} C.~O.,  {Zlochower} Y.,    {Merritt} D.,  2007,
  Physical Review Letters, 98, 231102

\bibitem[\protect\citeauthoryear{{Dotti}, {Colpi} \& {Haardt}}{{Dotti}
  et~al.}{2006}]{Dottietal2006}
{Dotti} M.,  {Colpi} M.,    {Haardt} F.,  2006, \mnras, 367, 103

\bibitem[\protect\citeauthoryear{{Dotti}, {Colpi}, {Haardt} \& {Mayer}}{{Dotti}
  et~al.}{2007}]{Dotti2007}
{Dotti} M.,  {Colpi} M.,  {Haardt} F.,    {Mayer} L.,  2007, MNRAS, 379, 956

\bibitem[\protect\citeauthoryear{{Dotti}, {Ruszkowski}, {Paredi}, {Colpi},
  {Volonteri} \& {Haardt}}{{Dotti} et~al.}{2009}]{Dotti2009a}
{Dotti} M.,  {Ruszkowski} M.,  {Paredi} L.,  {Colpi} M.,  {Volonteri} M.,
  {Haardt} F.,  2009, \mnras, 396, 1640

\bibitem[\protect\citeauthoryear{{Dotti}, {Volonteri}, {Perego}, {Colpi},
  {Ruszkowski} \& {Haardt}}{{Dotti} et~al.}{2010}]{Dotti2009b}
{Dotti} M.,  {Volonteri} M.,  {Perego} A.,  {Colpi} M.,  {Ruszkowski} M.,
  {Haardt} F.,  2010, \mnras, 402, 682

\bibitem[\protect\citeauthoryear{{Downes} \& {Solomon}}{{Downes} \&
  {Solomon}}{1998}]{Downes1998}
{Downes} D.,  {Solomon} P.~M.,  1998, ApJ, 507, 615

\bibitem[\protect\citeauthoryear{{Gammie}, {Shapiro} \& {McKinney}}{{Gammie}
  et~al.}{2004}]{Gammie2004}
{Gammie} C.~F.,  {Shapiro} S.~L.,    {McKinney} J.~C.,  2004, ApJ, 602, 312

\bibitem[\protect\citeauthoryear{{Gonz{\'a}lez}, {Sperhake}, {Br{\"u}gmann},
  {Hannam} \& {Husa}}{{Gonz{\'a}lez} et~al.}{2007}]{Gonzalez2007}
{Gonz{\'a}lez} J.~A.,  {Sperhake} U.,  {Br{\"u}gmann} B.,  {Hannam} M.,
  {Husa} S.,  2007, Physical Review Letters, 98, 091101

\bibitem[\protect\citeauthoryear{{Herrmann}, {Hinder}, {Shoemaker}, {Laguna} \&
  {Matzner}}{{Herrmann} et~al.}{2007}]{Herrmann2007}
{Herrmann} F.,  {Hinder} I.,  {Shoemaker} D.,  {Laguna} P.,    {Matzner} R.~A.,
   2007, ApJ, 661, 430

\bibitem[\protect\citeauthoryear{{Hills}}{{Hills}}{1975}]{Hills1975}
{Hills} J.~G.,  1975, \nat, 254, 295

\bibitem[\protect\citeauthoryear{{Ho}, {Filippenko} \& {Sargent}}{{Ho}
  et~al.}{1997}]{Ho1997}
{Ho} L.~C.,  {Filippenko} A.~V.,    {Sargent} W.~L.~W.,  1997, \apj, 487, 591

\bibitem[\protect\citeauthoryear{{Hopkins} \& {Hernquist}}{{Hopkins} \&
  {Hernquist}}{2006}]{HopkinsHernquist2006}
{Hopkins} P.~F.,  {Hernquist} L.,  2006, \apjs, 166, 1

\bibitem[\protect\citeauthoryear{{Hughes} \& {Blandford}}{{Hughes} \&
  {Blandford}}{2003}]{HB2003}
{Hughes} S.~A.,  {Blandford} R.~D.,  2003, \apjl, 585, L101

\bibitem[\protect\citeauthoryear{{Kharb}, {O'Dea}, {Baum}, {Colbert} \&
  {Xu}}{{Kharb} et~al.}{2006}]{Kharb2006}
{Kharb} P.,  {O'Dea} C.~P.,  {Baum} S.~A.,  {Colbert} E.~J.~M.,    {Xu} C.,
  2006, \apj, 652, 177

\bibitem[\protect\citeauthoryear{{King} \& {Pringle}}{{King} \&
  {Pringle}}{2006}]{King2006}
{King} A.~R.,  {Pringle} J.~E.,  2006, MNRAS, 373, L90

\bibitem[\protect\citeauthoryear{{K{\"o}rding}, {Jester} \&
  {Fender}}{{K{\"o}rding} et~al.}{2006}]{Kording2006}
{K{\"o}rding} E.~G.,  {Jester} S.,    {Fender} R.,  2006, MNRAS, 372, 1366

\bibitem[\protect\citeauthoryear{{Mayer}, {Kazantzidis}, {Madau}, {Colpi},
  {Quinn} \& {Wadsley}}{{Mayer} et~al.}{2007}]{Mayeretal2006}
{Mayer} L.,  {Kazantzidis} S.,  {Madau} P.,  {Colpi} M.,  {Quinn} T.,
  {Wadsley} J.,  2007, Science, 316, 1874

\bibitem[\protect\citeauthoryear{{Mayer}, {Kazantzidis}, {Mastropietro} \&
  {Wadsley}}{{Mayer} et~al.}{2007}]{Mayer2007}
{Mayer} L.,  {Kazantzidis} S.,  {Mastropietro} C.,    {Wadsley} J.,  2007,
  \nat, 445, 738

\bibitem[\protect\citeauthoryear{{Merloni} \& {Heinz}}{{Merloni} \&
  {Heinz}}{2008}]{Merloni2008}
{Merloni} A.,  {Heinz} S.,  2008, MNRAS, 388, 1011

\bibitem[\protect\citeauthoryear{{Milosavljevi{\'c}}, {Merritt} \&
  {Ho}}{{Milosavljevi{\'c}} et~al.}{2006}]{Milosavljevic2006}
{Milosavljevi{\'c}} M.,  {Merritt} D.,    {Ho} L.~C.,  2006, \apj, 652, 120

\bibitem[\protect\citeauthoryear{{Miniutti}, {Panessa}, {de Rosa}, {Fabian},
  {Malizia}, {Molina}, {Miller} \& {Vaughan}}{{Miniutti}
  et~al.}{2009}]{Miniutti2009}
{Miniutti} G.,  {Panessa} F.,  {de Rosa} A.,  {Fabian} A.~C.,  {Malizia} A.,
  {Molina} M.,  {Miller} J.~M.,    {Vaughan} S.,  2009, \mnras, 398, 255

\bibitem[\protect\citeauthoryear{{Perets}, {Hopman} \& {Alexander}}{{Perets}
  et~al.}{2007}]{Perets2007}
{Perets} H.~B.,  {Hopman} C.,    {Alexander} T.,  2007, \apj, 656, 709

\bibitem[\protect\citeauthoryear{{Rees}}{{Rees}}{1988}]{Rees1988}
{Rees} M.~J.,  1988, \nat, 333, 523

\bibitem[\protect\citeauthoryear{{Schmoll}, {Miller}, {Volonteri}, {Cackett},
  {Reynolds}, {Fabian}, {Brenneman}, {Miniutti} \& {Gallo}}{{Schmoll}
  et~al.}{2009}]{Schmoll2009}
{Schmoll} S.,  {Miller} J.~M.,  {Volonteri} M.,  {Cackett} E.,  {Reynolds}
  C.~S.,  {Fabian} A.~C.,  {Brenneman} L.~W.,  {Miniutti} G.,    {Gallo} L.~C.,
   2009, \apj, 703, 2171

\bibitem[\protect\citeauthoryear{{Soltan}}{{Soltan}}{1982}]{Soltan1982}
{Soltan} A.,  1982, \mnras, 200, 115

\bibitem[\protect\citeauthoryear{{Spaans} \& {Silk}}{{Spaans} \&
  {Silk}}{2000}]{Spaans00}
{Spaans} M.,  {Silk} J.,  2000, ApJ, 538, 115

\bibitem[\protect\citeauthoryear{{Thorne}}{{Thorne}}{1974}]{Thorne1974}
{Thorne} K.~S.,  1974, \apj, 191, 507

\bibitem[\protect\citeauthoryear{{Ulvestad} \& {Ho}}{{Ulvestad} \&
  {Ho}}{2001}]{Ulvestad2001}
{Ulvestad} J.~S.,  {Ho} L.~C.,  2001, ApJL, 562, L133

\bibitem[\protect\citeauthoryear{{Volonteri}, {Madau}, {Quataert} \&
  {Rees}}{{Volonteri} et~al.}{2005}]{Volonterietal2005}
{Volonteri} M.,  {Madau} P.,  {Quataert} E.,    {Rees} M.~J.,  2005, {ApJ},
  620, 69

\bibitem[\protect\citeauthoryear{{Volonteri}, {Sikora} \& {Lasota}}{{Volonteri}
  et~al.}{2007}]{VSL07}
{Volonteri} M.,  {Sikora} M.,    {Lasota} J.-P.,  2007, ApJ, 667, 704

\bibitem[\protect\citeauthoryear{{Wilson} \& {Colbert}}{{Wilson} \&
  {Colbert}}{1995}]{Wilson1995}
{Wilson} A.~S.,  {Colbert} E.~J.~M.,  1995, \apj, 438, 62

\end{thebibliography}

\end{document}